\pdfoutput=1

\documentclass[sigconf,screen]{acmart}

\usepackage{url}
\usepackage{booktabs} 
\usepackage{makecell} 
\usepackage{float} 
\usepackage{listings} 
\usepackage{enumitem}
  
\newcommand{\pit}{PITest}
\newcommand{\descartes}{Descartes}

\newcommand{\gregor}{Gregor}

\renewcommand\footnotetextcopyrightpermission[1]{} 
\begin{document}
\title{Descartes: A PITest Engine to Detect Pseudo-Tested Methods}
\subtitle{Tool Demonstration}

\author{Oscar Luis Vera-P\'erez}
\affiliation{
  \institution{Inria Rennes - Bretagne Atlantique}
  \streetaddress{}
  \city{Rennes}
  \country{France}
  \postcode{}
}
\email{oscar.vera-perez@inria.fr}

\author{Martin Monperrus}
\affiliation{
  \institution{KTH Royal Institute of Technology}
  \streetaddress{}
  \city{Stockholm}
  \country{Sweden}
  \postcode{}
}
\email{martin.monperrus@csc.kth.se}

\author{Benoit Baudry}
\affiliation{
  \institution{KTH Royal Institute of Technology}
  \streetaddress{}
  \city{Stockholm}
  \country{Sweden}
  \postcode{}
}
\email{baudry@kth.se}

\begin{abstract}
\descartes{} is a tool that implements extreme mutation operators and aims at finding pseudo-tested methods in Java projects. It leverages the efficient transformation and runtime features of \pit{}.  The demonstration compares \descartes{} with \gregor{}, the default mutation engine provided by \pit{}, in a set of real open source projects. It considers the execution time, number of mutants created and the relationship between the mutation scores produced by both engines. It provides some insights on the main features exposed by \descartes{}. 

\end{abstract}
  
\keywords{pseudo-tested methods, extreme mutation, mutation testing, software testing, \pit{}}

\maketitle

\section{Introduction}

Mutation analysis or mutation testing \cite{demillo_program_1979} evaluates the fault detection capabilities of a test suite. It does so by inserting artificial bugs in the form of subtle code changes. Then, it verifies if the test suite is able to detect those changes. The usual outcome from this analysis is the mutation score, that is, the ratio of planted faults (mutants) that has been detected to the total of mutants created.

Niedermayr and colleagues \cite{niedermayr_will_2016} recently introduced extreme mutation analysis. It is an alternative to traditional mutation that performs more coarse-grained transformations by eliminating, at once, all side effects of a method. For a $void$ method this approach removes all instructions from its body. If the method is not $void$, then the body is replaced by a single $return$ instruction with a predefined value. Listing \ref{list:method} shows a simple Java method and Listing \ref{list:mutation} shows two variants or mutants that could be created for this method using extreme mutation, in this case with constants $0$ and $1$. Besides removing all side effects, the technique ensures that the mutated method will always return the same value.

Extreme mutation addresses two challenges of the traditional approach. It creates much less mutants and can automatically avoid most transformations that could be equivalent to the original code. These two aspects are usually quoted as drawbacks that prevent the wide use of mutation testing in practice \cite{madeyski_overcoming_2014,mozucha_is_2016}. Another benefit of this approach is that it operates at the method level which eases the understanding of the underlying testing problem. In addition to the mutation score, extreme mutation pinpoints a list of worst tested methods. In particular, the technique higlights methods executed by the test suite but where no extreme mutant is detected while running the tests. These methods are labeled as pseudo-tested in the work of Niedermayr et. al.\cite{niedermayr_will_2016}.

In this demonstration, we present \descartes{}, an extreme mutation engine for \pit{} \cite{coles_pit_2016}, a state-of-the-art mutation testing tool for Java projects. \pit{} is a popular tool that works with all major build systems: Ant, Gradle, Maven and can handle JUnit and TestNG test suites. \descartes{} brings a set of extreme mutation operators to \pit{} and discovers pseudo-tested methods. We also compare the result provided by \descartes{} with the outcome of \gregor{}, the default mutation engine for \pit{}. Our goal is to determine if extreme mutation can be used as a viable trade-off between code coverage, which assesses only test inputs, and traditional mutation analysis, which also addresses the oracles but at a very high cost. This is a novel contribution with respect to the work of Niedermayr et al whose focus is on checking whether code coverage is a good indicator of test quality when discerning between system and unit tests.

\begin{lstlisting}[caption=A simple Java method., captionpos=b, label=list:method]
//Original method
public static long factorial(int n) {
    if(n==0) return 0;
    long result = 1;
    for(int i = 2; i <= n; i++)
        result *= i;
    return result;
}
\end{lstlisting}

\begin{lstlisting}[caption=Two mutans created with extreme mutation., captionpos=b, label=list:mutation]
//Extreme mutant 1
public static long factorial(int n) { return 0; }

//Extreme mutant 2
public static long factorial(int n) { return 1; }
\end{lstlisting}

\section{An overview of \descartes{}}

\descartes{} is a tool to automatically detect pseudo-tested methods in Java programs tested with JUnit test suites. This detection relies on extreme mutation analysis. We implement this analysis as a mutation engine for \pit{}. In \pit{}'s jargon, a mutation engine is a plugin that handles the discovery and creation of mutants. Such a plugin should also manage a set of mutation operators, which are models of the transformations to be performed.

Our extreme mutation engine provides a set of configurable mutation operators. A mutation operator is configured by specifying the literal value it should use to modify the method. \descartes{} supports literals of all Java primitive types, $String$, the $null$ value and has two special operators: one to target $void$ methods and another to return an empty array where possible. The engine does not mutate constructors.

\begin{figure*}
	\includegraphics[scale=.3]{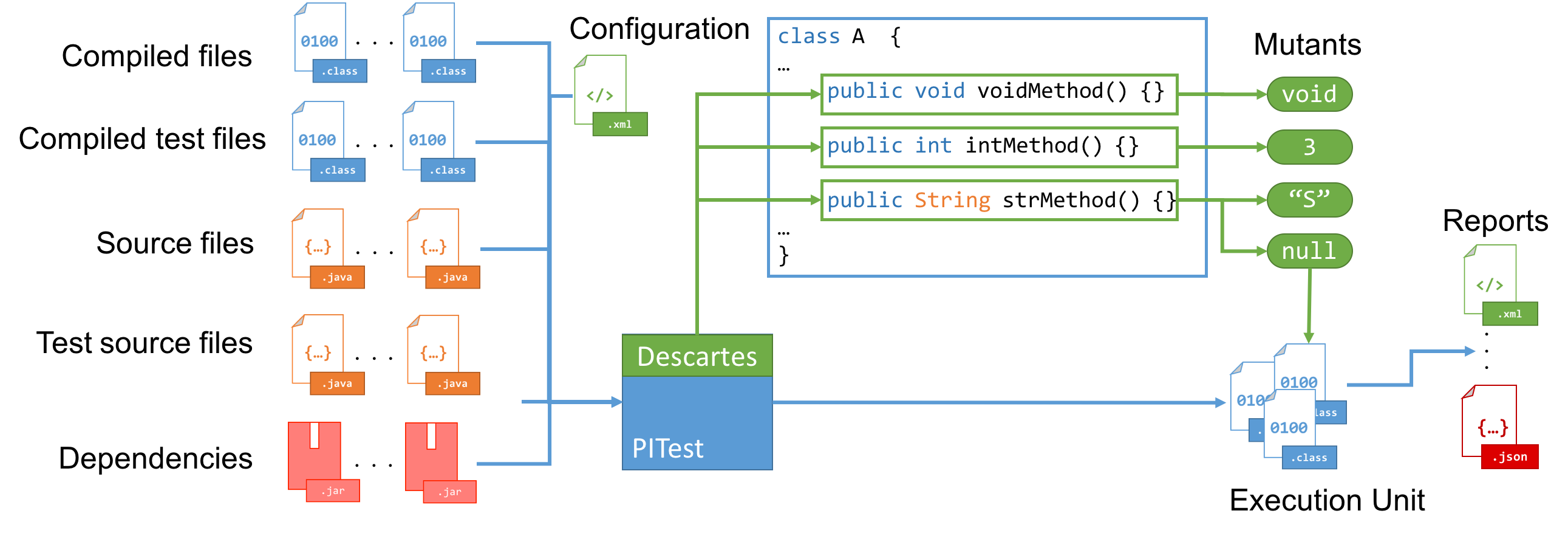}
	\caption{Interconnection between \pit{} and \descartes{}.}
	\label{fig:pitest-descartes}
\end{figure*}

Figure \ref{fig:pitest-descartes} illustrates the interaction between \pit{} and \descartes{}. \pit{} handles the inspection of the target project to discover all dependencies, creates execution units composed by the mutants and the tests to be executed, and ultimately runs the test cases. The mutation engine leverages all these functionalities and handles mutant discovery and creation.

Relying on the infrastructure and architecture of \pit{} allowed us to speed up the development of \descartes{} and its adoption in production. So far, the biggest challenges we have faced have been: 1) The lack of documentation describing how to create mutation engines for \pit{}; 2) The design and implementation of meaningful mutation operator abstractions and their interaction with the rest of the \pit{} framework; 3) Maintaining the engine up to date with the regular changes and releases of \pit{}; 4) Making the tool useful to developers. To overcome this last challenge we have augmented \descartes{} with custom reporting capabilities and functionalities to reduce the number of potential false positives, for example, we provide method filters based on the method structure rather than its signature.

To the best of our knowledge, \descartes{} is the only available alternative to the default engine provided by \pit{}. Our project could be used as an additional supporting material for those who are willing to create their own extensions.

\section{\descartes{} VS \gregor{}}

\gregor{} is the default mutation engine for \pit{}. It provides most traditional mutation operators \footnote{The full list is available here: \url{http://pitest.org/quickstart/mutators/}}. These operators work at the instruction level. Listing \ref{list:pitop} shows examples of the transformations that can be produced by \gregor{} over the method exposed in Listing \ref{list:method}. The first variant of the method, shown in line \ref{ln:gm1} negates the condition in line \ref{ln:condition}. The second variant, shown in line \ref{ln:gm2}, modifies the return value by adding $1$ in line \ref{ln:return}.

\begin{lstlisting}[caption=Examples of mutants produced by Gregor., label=list:pitop]
    //Mutant 1. Changes == by !=
    public static long factorial(int n) { %*\label{ln:gm1}*)
        if(n!=0) return 0; %*\label{ln:condition}*)
        long result = 1;
        for(int i = 2; i <= n; i++)
            result *= i;
        return result;
    }

    //Gregor mutant 2. Changes the result value by adding 1
    public static long factorial(int n) { %*\label{ln:gm2}*)
        if(n==0) return 0;
        long result = 1;
        for(int i = 2; i <= n; i++)
            result *= i;
        return result + 1; %*\label{ln:return}*)
    }
    \end{lstlisting}

We compare the execution of \gregor{} and \descartes{} in a selection of Java projects. These are all projects that use Maven as main build system, JUnit as main testing framework and are available form a version control hosting service, mostly Github.

Table \ref{tab:comparison} shows the metrics recorded for the comparison. For each mutation engine the table shows the execution time and number of mutants created. The ``\textit{Covered}'' columns show the number of mutants actually executed by the test suite and planted in methods that were mutated by both engines. This distinction removes from the comparison mutants that \gregor{} may create in methods not analyzed by \descartes{}, and vice versa. For example, mutants created in constructors are left out. The ``\textit{Killed}'' columns contain the number of mutants from the respective ``\textit{Covered}'' column that were detected (killed) by the test suite. The ``\textit{Score}'' columns show the corresponding mutation score, that is the ratio of ``\textit{Killed}'' to ``\textit{Covered}''.

\begin{table}[H]
	\caption{Extreme mutation operators used in the comparison.}
	\label{tab:desops}
	\begin{tabular}{ll}
		\toprule
		Method type         & Transformations        \\
        \midrule
        void                & Empties the method     \\
		Reference types     & Returns null           \\
		boolean             & Returns true or false  \\
		byte,short,int,long & Returns $0$ or $1$     \\
		float,double        & Returns $0.0$ or $0.1$ \\
		char                & Returns ` ' or `A'     \\
		String              & Returns ``'' or ``A''  \\
		T[]                 & Returns new T[]\{\}    \\
		\bottomrule
	\end{tabular}
\end{table}

For \gregor{}, all standard mutation operators were used. \descartes{} used the same mutation operators as Niedermayrs et. al. \cite{niedermayr_will_2016} plus two additional transformations, one to return $null$ for reference types and another to return an empty array. The full list of extreme mutation operators is shown in table \ref{tab:desops}.

\begin{table*}
    \caption{List of projects used to compare both engines, the execution time for the analysis, the number of mutants created, mutants covered and placed in methods targeted by both tools, mutants killed and the mutation score.}
    \label{tab:comparison}
\begin{tabular}{l|r|rrr|r||r|rrr|r}
\toprule

                    & \multicolumn{5}{c}{Descartes}   & \multicolumn{5}{c}{Gregor} \\

Project                      & Time    & Created &  Covered &  Killed &  Score & Time & Created &  Covered &  Killed  &  Score \\
\midrule
AuthZForce PDP Core          & 0:08:00 &     626 &  378 &   358 & 94.71 &  1:23:50 &    7296 &  3536 &  3188 & 90.16 \\
Amazon Web Services SDK      & 1:32:23 &  161758 & 3090 &  2732 & 88.41 &  6:11:22 & 2141689 & 17406 & 13536 & 77.77 \\
Apache Commons CLI           & 0:00:13 &     271 &  256 &   246 & 96.09 &  0:01:26 &    2560 &  2455 &  2183 & 88.92 \\
Apache Commons Codec         & 0:02:02 &     979 &  912 &   875 & 95.94 &  0:07:57 &    9233 &  8687 &  7765 & 89.39 \\
Apache Commons Collections   & 0:01:41 &    3558 & 1556 &  1463 & 94.02 &  0:05:41 &   20394 &  8144 &  7073 & 86.85 \\
Apache Commons IO            & 0:02:16 &    1164 & 1035 &   968 & 93.53 &  0:12:48 &    8809 &  7633 &  6500 & 85.16 \\
Apache Commons Lang          & 0:02:07 &    3872 & 3261 &  3135 & 96.14 &  0:21:02 &   30361 & 25431 & 22120 & 86.98 \\
Apache Flink                 & 0:14:04 &    4935 & 2781 &  2373 & 85.33 &  2:29:45 &   43619 & 21350 & 16647 & 77.97 \\
Google Gson                  & 0:01:08 &     848 &  657 &   617 & 93.91 &  0:05:34 &    7353 &  6179 &  5079 & 82.20 \\
Jaxen XPath Engine           & 0:01:31 &    1252 &  953 &   921 & 96.64 &  0:24:40 &   12210 &  9002 &  6041 & 67.11 \\
JFreeChart                   & 0:05:48 &    7210 & 4686 &  3775 & 80.56 &  0:41:28 &   89592 & 47305 & 28080 & 59.36 \\
Java Git                     & 1:30:08 &    7152 & 5007 &  4507 & 90.01 & 16:02:03 &   78316 & 54441 & 40756 & 74.86 \\
Joda-Time                    & 0:03:39 &    4525 & 3996 &  3827 & 95.77 &  0:16:32 &   31233 & 26443 & 21911 & 82.86 \\
JOpt Simple                  & 0:00:37 &     412 &  397 &   379 & 95.47 &  0:01:36 &    2271 &  2136 &  2000 & 93.63 \\
jsoup                        & 0:02:43 &    1566 & 1248 &  1197 & 95.91 &  0:12:49 &   14054 & 11092 &  8771 & 79.08 \\
SAT4J Core                   & 0:53:09 &    2304 &  804 &   617 & 76.74 & 10:55:50 &   17163 &  7945 &  5489 & 69.09 \\
Apache PdfBox                & 0:44:07 &    7559 & 3185 &  2548 & 80.00 &  6:20:25 &   79763 & 32753 & 20226 & 61.75 \\
SCIFIO                       & 0:24:14 &    3627 & 1235 &  1158 & 93.77 &  3:12:11 &   62768 & 19615 &  9496 & 48.41 \\
Spoon                        & 2:24:55 &    4713 & 3452 &  3171 & 91.86 & 56:47:57 &   43916 & 34694 & 27519 & 79.32 \\
Urban Airship Client Library & 0:07:25 &    3082 & 2362 &  2242 & 94.92 &  0:11:31 &   17345 & 11015 &  8956 & 81.31 \\
XWiki Rendering Engine       & 0:10:56 &    5534 & 3099 &  2594 & 83.70 &  2:07:19 &  112605 & 50472 & 26292 & 52.09 \\
\bottomrule
\end{tabular}
\end{table*}

One can observe that \descartes{} creates much less mutants than \gregor{} which is reflected in the difference between the times to execute the analysis of each engine. In all cases, \descartes{} completed the task in much less time. Some interesting contrasts in this matter come from projects like \texttt{Spoon} where \descartes{} took a little less than two hours and a half while \gregor{} took more than 56 hours, \texttt{Java Git} with one hour and a half for extreme mutation and 16 hours for \gregor{} and \texttt{Jaxen XPath Engine} with less than two minutes against nearly 25 minutes. While the number of mutants created and covered affects the execution time, the tests themselves play an important role as they can involve heavy computation. Take, for example, the difference between \texttt{Apache Commons Lang} and \texttt{SCIFIO} with similar numbers of mutants and very different execution times.

\begin{figure}
	\includegraphics[scale=.4]{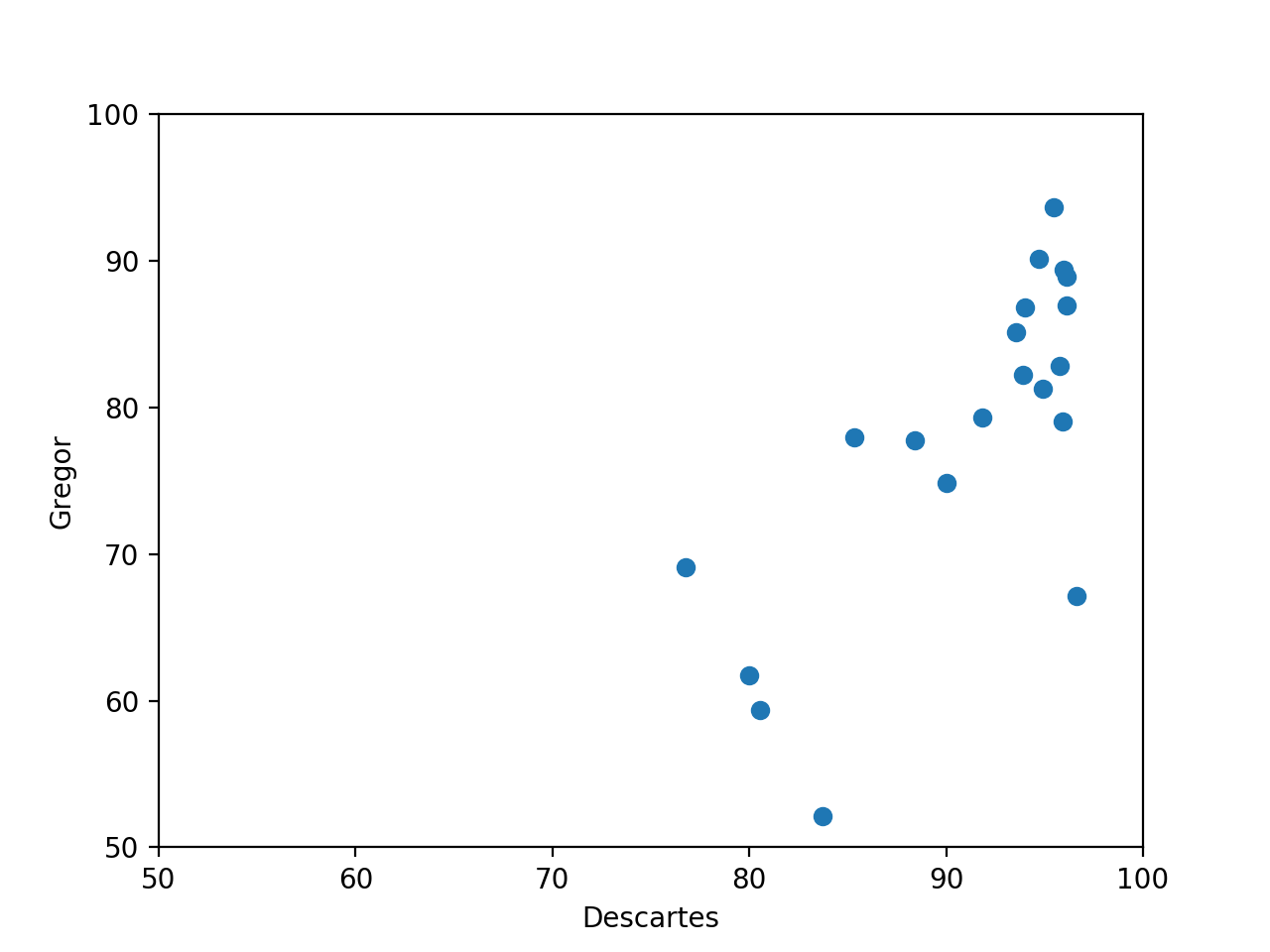}
	\caption{Visual correlation between scores}
	\label{fig:score-correlation}
\end{figure}

As for the scores, one can notice that there is a certain correlation between the values obtained by both engines. Figure \ref{fig:score-correlation} shows a scatter plot, in which each point represents a project. The coordinates for each point are given by the scores, the \textit{x} axis corresponds to the score from \descartes{} while the \textit{y} represents the score from \gregor{}. The figure corroborates the tendency for a positive monotonic correlation between both scores, which means that, if the score with \gregor{} is high, it is more likely that the mutation score with \descartes{} will be also high. The Spearman correlation coefficient results in 0.6 for the projects studied with a \textit{p-value} of 0.003, which indeed indicates that there is a moderate positive correlation. Anyways, there are cases such as \texttt{SCIFIO} and \texttt{XWiki Rendering Engine} which produce a medium to low mutation score with \gregor{} and scores above 83\% with \descartes{}.

\section{Pseudo-tested methods}

The results of extreme mutation are not limited to produce a score for a given project. The proposal of Niedermayr et. al. \cite{niedermayr_will_2016} classifies methods according to the extreme mutant detection. A method is said to be \textbf{pseudo-tested} if it is covered by the test suite but no related extreme mutant is killed. These methods are the worst tested in the code base. Extreme mutation provides a framework to detect such methods more efficiently than traditional mutation testing.

Listing \ref{list:pseudo} shows a method belonging to one of the projects included in table \ref{tab:comparison}. It was found to be pseudo-tested by \descartes{}. Only two extreme mutations are required to detect that the value of this method is not correctly verified by the test suite, if verified at all, while \gregor{} created 45 mutants. This is an example of the utility of extreme mutation.

\begin{lstlisting}[caption=Real example of a pseudo-tested method.,label=list:pseudo, float, floatplacement=H]
public static boolean isValidXmlChar(int ch) {
    return (ch == 0x9)
    || (ch == 0xA)
    || (ch == 0xD)
    || (ch >= 0x20 && ch <= 0xD7FF)
    || (ch >= 0xE000 && ch <= 0xFFFD)
    || (ch >= 0x10000 && ch <= 0x10FFFF);
}
\end{lstlisting}

Nevertheless, the result of \descartes{} is coarse-grained. Methods where extreme mutants are killed are not exempt from having testing issues. Listing \ref{list:tested} shows a real example of a method where all extreme mutants were detected but \gregor{} created mutants that survived the analysis. In particular one of the traditional mutation operators changed the value of $Long.MIN\_VALUE$ in line \ref{line:condition}. The modification was unnoticed by the test suite, which indicates that the corner case is not being tested. This level of detail can not be reached with the use of extreme mutation alone.

For a deep analysis regarding the utility in practice of \descartes{} in the search of pseudo-tested methods we invite the reader to check our work on the matter \cite{vera_comprehensive_2018}. There, we analyze whether these methods are valid hints to improve existing test cases and we provide a set of testing issues found with the help of \descartes{} in real and well tested open-source projects.

\begin{lstlisting}[caption=Example of a non pseudo-tested method., captionpos=b, label=list:tested]
public long subtract(long instant, long value) {
  if (value == Long.MIN_VALUE) %*\label{line:condition}*)
      throw new ArithmeticException(...);
  return add(instant, -value);
}
\end{lstlisting}

\section{Demonstration scope}

The demonstration will be directed to researchers and developers who wish to experiments with traditional and extreme mutation. It will be focused on the practical comparison of both mutation approaches. We will discuss how to interpret the results given by \descartes{} and how practitioners can use these results to enhance their test suites. We will show examples of real testing faults found with the use of the extreme mutation engine. The demo will also showcase the integration of \descartes{} and the latest check Github API \footnote{https://developer.github.com/v3/checks/} to discover pseudo-tested methods in commits and pull requests.

\section{Supporting Materials}

All materials related to the tool are available online. Here we provide a list with the main resources:

\begin{itemize}[leftmargin=0.5cm]
    \item \textbf{\descartes{} code repository}: Main code repository hosted in Github. It contains the code, documentation and instructions to build the tool. 
    
    \url{https://github.com/STAMP-project/pitest-descartes}
    
    \item \textbf{Experimental data}: Consists in a set of files with the output obtained from both mutation engines as well as data concerning the studied projects.
    
    \url{https://figshare.com/articles/data/6343280}
    
    \item \textbf{Experimental material}: 
    Github repository with additional experimental data and scripts to support the analysis and comparison of both mutation engines.
    
    \url{https://github.com/STAMP-project/descartes-experiments} 

    \item \textbf{Maven Central artifacts}: Compiled versions of Descartes are available for use from Maven Central. 
    
    \url{https://mvnrepository.com/artifact/eu.stamp-project/descartes}

    \item \textbf{Github Application repository}: Code of the prototype application to integrate \descartes{} in a Github repository. 

    \url{https://github.com/STAMP-project/descartes-github-app}
\end{itemize}

\section*{Acknowledgments}
This work has been supported by the EU Project STAMP ICT-16-10 No.731529.

\bibliographystyle{ACM-Reference-Format}
\bibliography{references}


\begin{thebibliography}{6}


\ifx \showCODEN    \undefined \def \showCODEN     #1{\unskip}     \fi
\ifx \showDOI      \undefined \def \showDOI       #1{#1}\fi
\ifx \showISBNx    \undefined \def \showISBNx     #1{\unskip}     \fi
\ifx \showISBNxiii \undefined \def \showISBNxiii  #1{\unskip}     \fi
\ifx \showISSN     \undefined \def \showISSN      #1{\unskip}     \fi
\ifx \showLCCN     \undefined \def \showLCCN      #1{\unskip}     \fi
\ifx \shownote     \undefined \def \shownote      #1{#1}          \fi
\ifx \showarticletitle \undefined \def \showarticletitle #1{#1}   \fi
\ifx \showURL      \undefined \def \showURL       {\relax}        \fi
\providecommand\bibfield[2]{#2}
\providecommand\bibinfo[2]{#2}
\providecommand\natexlab[1]{#1}
\providecommand\showeprint[2][]{arXiv:#2}

\bibitem[\protect\citeauthoryear{Coles, Laurent, Henard, Papadakis, and
  Ventresque}{Coles et~al\mbox{.}}{2016}]%
        {coles_pit_2016}
\bibfield{author}{\bibinfo{person}{Henry Coles}, \bibinfo{person}{Thomas
  Laurent}, \bibinfo{person}{Christopher Henard}, \bibinfo{person}{Mike
  Papadakis}, {and} \bibinfo{person}{Anthony Ventresque}.}
  \bibinfo{year}{2016}\natexlab{}.
\newblock \showarticletitle{{PIT}: {A} {Practical} {Mutation} {Testing} {Tool}
  for {Java} ({Demo})}. In \bibinfo{booktitle}{\emph{Proceedings of the 25th
  {International} {Symposium} on {Software} {Testing} and {Analysis}}}
  \emph{(\bibinfo{series}{{ISSTA} 2016})}. \bibinfo{publisher}{ACM},
  \bibinfo{address}{New York, NY, USA}, \bibinfo{pages}{449--452}.
\newblock
\showISBNx{978-1-4503-4390-9}
\urldef\tempurl%
\url{https://doi.org/10.1145/2931037.2948707}
\showDOI{\tempurl}


\bibitem[\protect\citeauthoryear{DeMillo, Lipton, and Sayward}{DeMillo
  et~al\mbox{.}}{1979}]%
        {demillo_program_1979}
\bibfield{author}{\bibinfo{person}{Richard~A. DeMillo},
  \bibinfo{person}{Richard~J. Lipton}, {and} \bibinfo{person}{Frederick~G.
  Sayward}.} \bibinfo{year}{1979}\natexlab{}.
\newblock \showarticletitle{Program mutation: A new approach to program
  testing}.
\newblock \bibinfo{journal}{\emph{Infotech State of the Art Report, Software
  Testing}} \bibinfo{volume}{2}, \bibinfo{number}{1979} (\bibinfo{year}{1979}),
  \bibinfo{pages}{107--126}.
\newblock


\bibitem[\protect\citeauthoryear{Madeyski, Orzeszyna, Torkar, and
  J\'ozala}{Madeyski et~al\mbox{.}}{2014}]%
        {madeyski_overcoming_2014}
\bibfield{author}{\bibinfo{person}{Lech Madeyski}, \bibinfo{person}{Wojciech
  Orzeszyna}, \bibinfo{person}{Richard Torkar}, {and} \bibinfo{person}{Mariusz
  J\'ozala}.} \bibinfo{year}{2014}\natexlab{}.
\newblock \showarticletitle{Overcoming the {Equivalent} {Mutant} {Problem}: {A}
  {Systematic} {Literature} {Review} and a {Comparative} {Experiment} of
  {Second} {Order} {Mutation}}.
\newblock \bibinfo{journal}{\emph{IEEE Transactions on Software Engineering}}
  \bibinfo{volume}{40}, \bibinfo{number}{1} (\bibinfo{date}{Jan.}
  \bibinfo{year}{2014}), \bibinfo{pages}{23--42}.
\newblock
\showISSN{0098-5589}
\urldef\tempurl%
\url{https://doi.org/10.1109/TSE.2013.44}
\showDOI{\tempurl}


\bibitem[\protect\citeauthoryear{Mo\v{z}ucha and Rossi}{Mo\v{z}ucha and
  Rossi}{2016}]%
        {mozucha_is_2016}
\bibfield{author}{\bibinfo{person}{Jakub Mo\v{z}ucha} {and}
  \bibinfo{person}{Bruno Rossi}.} \bibinfo{year}{2016}\natexlab{}.
\newblock \showarticletitle{Is {Mutation} {Testing} {Ready} to {Be} {Adopted}
  {Industry}-{Wide}?}. In \bibinfo{booktitle}{\emph{Product-{Focused}
  {Software} {Process} {Improvement}}} \emph{(\bibinfo{series}{Lecture {Notes}
  in {Computer} {Science}})}. \bibinfo{publisher}{Springer, Cham},
  \bibinfo{pages}{217--232}.
\newblock
\showISBNx{978-3-319-49093-9 978-3-319-49094-6}
\urldef\tempurl%
\url{https://doi.org/10.1007/978-3-319-49094-6_14}
\showDOI{\tempurl}


\bibitem[\protect\citeauthoryear{Niedermayr, Juergens, and Wagner}{Niedermayr
  et~al\mbox{.}}{2016}]%
        {niedermayr_will_2016}
\bibfield{author}{\bibinfo{person}{Rainer Niedermayr}, \bibinfo{person}{Elmar
  Juergens}, {and} \bibinfo{person}{Stefan Wagner}.}
  \bibinfo{year}{2016}\natexlab{}.
\newblock \showarticletitle{Will my tests tell me if {I} break this code?}. In
  \bibinfo{booktitle}{\emph{Proceedings of the {International} {Workshop} on
  {Continuous} {Software} {Evolution} and {Delivery}}}. \bibinfo{publisher}{ACM
  Press}, \bibinfo{address}{New York, NY, USA}, \bibinfo{pages}{23--29}.
\newblock
\showISBNx{978-1-4503-4157-8}
\urldef\tempurl%
\url{https://doi.org/10.1145/2896941.2896944}
\showDOI{\tempurl}


\bibitem[\protect\citeauthoryear{Vera-P\'erez, Danglot, Monperrus, and
  Baudry}{Vera-P\'erez et~al\mbox{.}}{2018}]%
        {vera_comprehensive_2018}
\bibfield{author}{\bibinfo{person}{Oscar~Luis Vera-P\'erez},
  \bibinfo{person}{Benjamin Danglot}, \bibinfo{person}{Martin Monperrus}, {and}
  \bibinfo{person}{Benoit Baudry}.} \bibinfo{year}{2018}\natexlab{}.
\newblock \showarticletitle{A {Comprehensive} {Study} of {Pseudo-tested}
  {Methods}}.
\newblock \bibinfo{journal}{\emph{arXiv:1807.05030 [cs]}} (\bibinfo{date}{July}
  \bibinfo{year}{2018}).
\newblock
\urldef\tempurl%
\url{http://arxiv.org/abs/1807.05030}
\showURL{%
\tempurl}
\newblock
\shownote{arXiv: 1807.05030.}


\end{thebibliography}

\end{document}